\documentclass[aps,prl,epsfig,graphics,twocolumn,floatfix,mathbbm,a4paper]{revtex4}

\usepackage{amsmath,amsfonts,amssymb,graphics,graphicx,epsfig,color,times,bbm}

\begin{document}
\bibliographystyle{apsrev}

\newcommand{\R}{\mathbbm{R}}
\newcommand{\rr}{\mathbbm{R}}
\newcommand{\nn}{\mathbbm{N}}
\newcommand{\cc}{\mathbbm{C}}
\newcommand{\ii}{\mathbbm{1}}

\newcommand{\id}{\mathbbm{1}}

\newcommand{\tr}{{\rm tr}}
\newcommand{\gr}[1]{\boldsymbol{#1}}
\newcommand{\be}{\begin{equation}}
\newcommand{\ee}{\end{equation}}
\newcommand{\bea}{\begin{eqnarray}}
\newcommand{\eea}{\end{eqnarray}}
\newcommand{\ket}[1]{|#1\rangle}
\newcommand{\bra}[1]{\langle#1|}
\newcommand{\avr}[1]{\langle#1\rangle}
\newcommand{\G}{{\cal G}}
\newcommand{\eq}[1]{Eq.~(\ref{#1})}
\newcommand{\ineq}[1]{Ineq.~(\ref{#1})}
\newcommand{\sirsection}[1]{\section{\large \sf \textbf{#1}}}
\newcommand{\sirsubsection}[1]{\subsection{\normalsize \sf \textbf{#1}}}
\newcommand{\ack}{\subsection*{\normalsize \sf \textbf{Acknowledgements}}}
\newcommand{\front}[5]{\title{\sf \textbf{\Large #1}}
\author{#2 \vspace*{.4cm}\\
\footnotesize #3}
\date{\footnotesize \sf \begin{quote}
\hspace*{.2cm}#4 \end{quote} #5} \maketitle}
\newcommand{\eg}{\emph{e.g.}~}

\newcommand{\proofend}{\hfill\fbox\\\medskip }


\newtheorem{theorem}{Theorem}
\newtheorem{proposition}{Proposition}

\newtheorem{lemma}{Proposition}

\newtheorem{definition}{Definition}
\newtheorem{corollary}{Corollary}

\newcommand{\proof}[1]{{\bf Proof.} #1 $\proofend$}

\title{Quantum phase transitions in matrix product systems}

\author{Michael M. Wolf$^{1}$, Gerardo Ortiz$^{2}$, F. Verstraete$^3$ and J. Ignacio Cirac$^{1}$}
\affiliation{$^{1}$ Max-Planck-Institute for Quantum Optics,
 Hans-Kopfermann-Str.\ 1, D-85748 Garching, Germany.\\$^{2}$ Los Alamos National Laboratory, Los Alamos, New Mexico 87545, USA.\\ $^3$ Institute for Quantum
Information, Caltech, Pasadena, US.}

\date{\today}


\begin{abstract}
We investigate quantum phase transitions (QPTs) in spin chain
systems characterized by local Hamiltonians with matrix product
ground states. We show how to theoretically engineer such QPT
points between states with predetermined properties. While some of
the characteristics of these transitions are familiar, like the
appearance of singularities in the thermodynamic limit, diverging
correlation length, and vanishing energy gap, others differ from
the standard paradigm: In particular, the ground state energy
remains analytic, and the entanglement entropy of a half-chain
stays finite. Examples demonstrate that these kinds of transitions
can occur at the triple point of `{\it conventional}' QPTs.
\end{abstract}

\maketitle



A considerable part of modern condensed matter physics is devoted to the study of
matter near zero temperature. In particular, zero-temperature quantum phase transitions (QPTs) \cite{Sac99}, as
observed in cuprate high-temperature superconductors and heavy fermion materials, have attracted enormous
attention. Although an adaptation of the classical Landau-Ginzburg theory successfully describes some of these
phase transitions, it is manifest that this concept is in general too narrow to cover all the fascinating aspects
of the quantum world \cite{SRIS01}. A complete and rigorous quantum mechanical description is, however, burdened
by the notorious complexity of quantum correlations in highly entangled many-body systems.

The fields of condensed matter and quantum information theory
study the behavior of quantum many-body systems by using
complementary methodologies. Whereas the typical starting point in
condensed matter theory is a Hamiltonian, from which states emerge
as ground states (GSs) or excitations, quantum information theory
deals primarily with quantum states, from which corresponding
Hamiltonians may be constructed. For spin chains this point of
view can be traced back to the seminal works on the AKLT model
\cite{AKLT88} and finitely correlated states \cite{FNW92}, and it
has recently been successfully resumed in various works on matrix
product states (MPS) \cite{MPS}. This led to new powerful
numerical algorithms \cite{VPC04,VGC04,ZV04} accompanied by a
better understanding of their efficiency \cite{VC05}, and new
insights in renormalization group transformations \cite{VCLRW05}
and sequential quantum generators \cite{SSVCW05}. Clearly, a
fruitful crossfertilization is emerging between these two fields.

The present work investigates QPTs in systems represented by MPS
by following the quantum information approach. It thus generalizes
the findings of \cite{FNW92,MPS2} which already indicated the
possibility of such transitions in MPS systems, and it enables us
to theoretically design QPT's in quasi-exactly solvable models. In
this way, we may engineer QPT points between phases whose
correlations or symmetries we choose a priori. The corresponding
orders can be of local type and/or of a more subtle hidden
non-local character. The main observation behind is that, for the
systems under consideration, a singularity in a $({\cal
D}-1)$-dimensional transfer operator leads to a QPT in the
corresponding ${\cal D}$-dimensional quantum system. Although
these findings hold for arbitrary $\cal D$, we focus on MPS in
${\cal D}=1$, where a general discussion is possible on full
analytic grounds.

In fact, \emph{every state}, in particular every GS, of a finite
system can be represented as a MPS \cite{FNW92,VPC04}. The power
of this representation ---and with it the power of Density Matrix
Renormalization Group (DMRG)--- stems from the fact that in many
cases a low-dimensional MPS already leads to a very good
approximation of the state \cite{VC05}. From such a
low-dimensional MPS one can in turn construct a parent Hamiltonian
from which it arises as an \emph{exact} GS. We will study the
dependence of correlation functions of such systems on a smoothly
varying parameter $g$ and show that singularities can appear,
which are reminiscent to those arising in known examples of QPTs.
They appear only in the thermodynamic limit and are accompanied by
diverging correlation lengths and vanishing energy gaps.

Some of the derived properties do, however, hardly fit within the
conventional picture of QPTs in ${\cal D}=1$ spin systems: First,
at the QPT point $g=g_c$ the GS energy density $e_0$ is analytic
(Fig. 1). Typically, a non-analyticity in $e_0(g)$ is used as the
defining property of a QPT \cite{Sac99}. We think, however, that
the use of the term QPT is well justified by the presence of a
non-analyticity of {\it any} observable quantity (e.g., two-point
correlations).  The second non-typical feature is the fact that
the entropy of a half-chain remains finite as $g\rightarrow g_c$,
which reflects the fact that MPS-QPTs cannot be described in terms
of conformal field theory. In fact, a lot of attention has
recently been devoted to the {\it entanglement entropy}
\cite{Latorre,area}, resulting in the observation that the
crossing of a QPT point typically coincides with the divergence of
this entropy. The discussion below shows, however, that this is
not the case for MPS QPTs in ${\cal D}=1$.
A third peculiarity is, that although the breaking of a discrete
symmetry can be engineered, MPS QPTs can occur without spontaneous
symmetry breaking since for $g\neq g_c$ the GS is typically
unique.

We will start by discussing the relevant properties of MPS and show how singularities can appear in the
thermodynamic limit (size of the chain $N \rightarrow \infty$). Then, following the idea of \cite{FNW92} we show
how the corresponding Hamiltonians can be constructed and discuss two examples, with local and non-local order,
in greater detail. For one of these we show that the MPS-QPT occurs exactly at the triple point of `{\it
conventional}' QPTs.

\emph{Singularities in MPS} --- Consider a MPS which is up to normalization given by \be\label{mps} |\psi\rangle
= \sum_{i_1...i_N=1}^d \tr[A_{i_1}...A_{i_N}]\ |i_1 ... i_N \rangle\;, \ee where $\{A_i\}$ is a set of $d$
$D\times D$ matrices, $d$ is the Hilbert space dimension corresponding to one site in the chain and $D$ is the
dimension of the bonds, when we think of the state in the valence bond picture. The state in Eq. (\ref{mps}) is
translational invariant on a ring of length $N$, it has reflection symmetry if $A_i=A_i^T$, permutation symmetry
if $[A_i,A_j]=0$ and time reversal symmetry if the $A_i$'s are real. Also other local symmetries (e.g., $SU(2)$,
$\mathbb{Z}_2$) can be enforced by imposing appropriate constraints on the $A_i$'s \cite{FNW92,Nac96}.

 In the following we will consider systems
where the matrices $A_i$ depend on a single real parameter $g$. It is important to note that if the $A_i$ depend
on $g$ in an analytic or continuous way, then so will its parent Hamiltonian which we construct below. Clearly,
we could consider more general cases with several parameters $g_1,g_2,\ldots$ .

Correlation functions for $m$ consecutive sites are given by \bea\langle S_1\ldots S_{m}\rangle &=&
\frac{\tr[E_{\ii}^{N-m}
 E_{S_1}\ldots E_{S_m}
]}{\tr[E_{\ii}^{N}]},\ \ \mbox{with}\label{npoint} \\
E_S &=& \sum_{i,j=1}^d \langle i |S|j\rangle A_j\otimes\bar{A_i}.\label{corr}\eea Here $S_i$ is any observable
acting on the $i^{\rm th}$ site and the bar denotes complex conjugation. For simplicity we will focus on the
generic case where the {\it transfer operator} $E_\ii$ is diagonalizable and non-degenerate for $g\neq g_c$.
Taking the thermodynamic limit ($N\rightarrow\infty$) only the right $\ket{r}$ and left $\ket{r}$ eigenvectors of
$E_{\ii}$'s largest eigenvalue $\nu_1$ survive in Eq.~(\ref{npoint}). With the normalization $\langle
l|r\rangle=1$ this leads to \be\label{rlnu} \langle S_1\ldots S_m\rangle = \langle l| E_{S_1}\ldots E_{S_{m}}
|r\rangle / \nu_1^m\;.\ee Hence, if for some $g=g_c$ there is a level crossing in the largest eigenvalues of
$E_{\ii}$, then there will typically be a discontinuity in the correlation functions (or their derivatives), even
though the $A_i$'s and with them the $E$'s are analytic in $g$. Needless to say that the same argumentation holds
for every observable with finite support.

A trivial example showing that discontinuities of any order $n$ are possible is given by ($D=d=2$)\be A_1=\left(
\begin{array}{cc}
  1 & 0\\
  0 & 1+g \\
\end{array}
\right),\quad A_2=\left(
\begin{array}{cc}
  g^n & 0 \\
  0 & 0 \\
\end{array}
\right). \ee Here, all derivatives $\partial_g^k \langle S_i S_{i+1}\rangle$ of order $k<n$ will be continuous at
$g=g_c=0$, whereas the $n^{\rm th}$-order derivative e.g. of $\langle \sigma^x  \rangle$ turns out to be
discontinuous.

Let us now discuss the properties of a general MPS in the vicinity
of a transition point $g_c$. The decay of two-point correlations
$\langle S_i S_{i+l}\rangle$ can be obtained from Eq.
(\ref{npoint}) by setting $E_{S_2}=\ldots=E_{S_{m-1}}=E_\ii$ with
$m=l+1$ and exploiting the Jordan decomposition of the transfer
operator. This leads to \be \langle S_i S_{i+l}\rangle - \langle
S_i\rangle \langle S_{i+l}\rangle
\sim\Big|\frac{\nu_2}{\nu_1}\Big|^{\;l-1}\;,\ee where $\nu_2$ is
the second largest eigenvalue of the transfer operator. As the
coupling strength approaches its QPT point value,  $g\rightarrow
g_c$, we get $|\nu_2|\rightarrow|\nu_1|$. Then, the correlation
length $\xi=1/\log\big|\nu_1/\nu_2\big|$ diverges and one obtains
long-range correlations at the transition point. Note that despite
the diverging correlation length there is no power-law decay at
the transition point (which can be different for ${\cal D}> 1 $
\cite{elsewhere}).

Since a lot of attention has recently been devoted to the relation
between criticality and the scaling of the entanglement entropy
\cite{Latorre} we give now an explicit formula for the latter. In
fact, in recent works QPT points seem to be intimately connected
with a logarithmic diverging behavior of the entropy of a block of
consecutive spins when considered as a function of the block size.
However, this is not the case for the class of QPT's discussed in
the present work. The entropy of an asymptotically large block of
a MPS can be calculated exactly by exploiting the freedom in the
$A_i$'s in order to fix the gauge $\sum_iA_iA_i^\dag =\ii$,
$\sum_iA_i^\dag \varrho A_i=\varrho$, where $\varrho$ is a density
matrix acting on $\mathbb{C}^D$. By the renormalization group
arguments of \cite{VCLRW05,FNW92} the spectrum of a large block
converges to the spectrum of $\varrho^{\otimes 2}$ such that the
entropy becomes $2 S(\varrho)=-2\tr{\varrho\log_2\varrho}$. In
particular, the entropy never exceeds $2\log D$, irrespective of
how close the system is to a QPT point. This immediately implies
that states exhibiting a diverging growth of the entanglement
entropy cannot be described by MPS with finite $D$ (explaining the
difficulties of DMRG for critical systems). Moreover, following
\cite{CC} the absence of a logarithmic divergence implies that
MPS-QPTs cannot be described in terms of conformal field theory
(as already indicated by the absence of power-law decaying
correlations). \vspace*{3pt}

\emph{Higher dimensions} --- Although our main focus is on one-dimensional quantum systems (${\cal D}=1$) we
briefly discuss in this interlude how to extend the presented ideas to higher dimensions.

Note that in the above ${\cal D}=1$ case the QPT is traced back to
a level crossing in the operator $E_\ii$, which acts only on a
single site, i.e., in dimension ${\cal D}-1$. However, if the
transfer operator itself has a spatial substructure, we are led to
a QPT in a higher dimensional system. To be more specific consider
a ${\cal D}$-dimensional cubic lattice of size
$N_1\times\ldots\times N_{\cal D}$ with periodic boundary
conditions. A generalization of (\ref{mps}), the so-called
\emph{projected entangled pair state} (PEPS \cite{PEPS}), is then
obtained by replacing the bi-linear forms $A_i$ by tensors of
order $2{\cal D}$, i.e., $A_i:(\mathbb{C}^{D})^{\otimes 2{\cal
D}}\rightarrow\mathbb{C}$ and the matrix product by a tensor
contraction according to the edges of the lattice. As we can
interpret the $\cal D$-dimensional lattice as a chain of ${\cal
D}-1$ dimensional systems we can introduce a transfer operator
$E_{\ii}'$ for this {\it chain} by contracting all
$\prod_{i=2}^{\cal D} N_i$ operators $E_{\ii}$ on a ${\cal D}-1$
dimensional sub-lattice. In this way we are back to the
one-dimensional scenario described above. That is, if we take
$N_1\rightarrow\infty$ then a level-crossing in the largest
eigenvalue of $E_\ii'$ can give rise to a QPT. Note that this way
of constructing QPT points resembles the classical transfer matrix
method discussed for instance in \cite{Kac}. The problem of
calculating the largest eigenvalue of $E_\ii'$ can numerically be
tackled by DMRG (PEPS) algorithms for ${\cal D}=2$ (${\cal D}\geq
3$). Particular instances of such transitions were discussed in
\cite{2D}, where power-law decaying correlations
 could be determined using Monte-Carlo methods. Note that by
 construction all \emph{critical} ground states obtained in this way obey an
\emph{area law} for the entanglement entropy \cite{area}. A method
for deriving analytic results for particular higher dimensional
instances will be provided elsewhere \cite{elsewhere}.
\vspace{3pt}

\begin{figure}[t]
\begin{center}
\epsfig{file=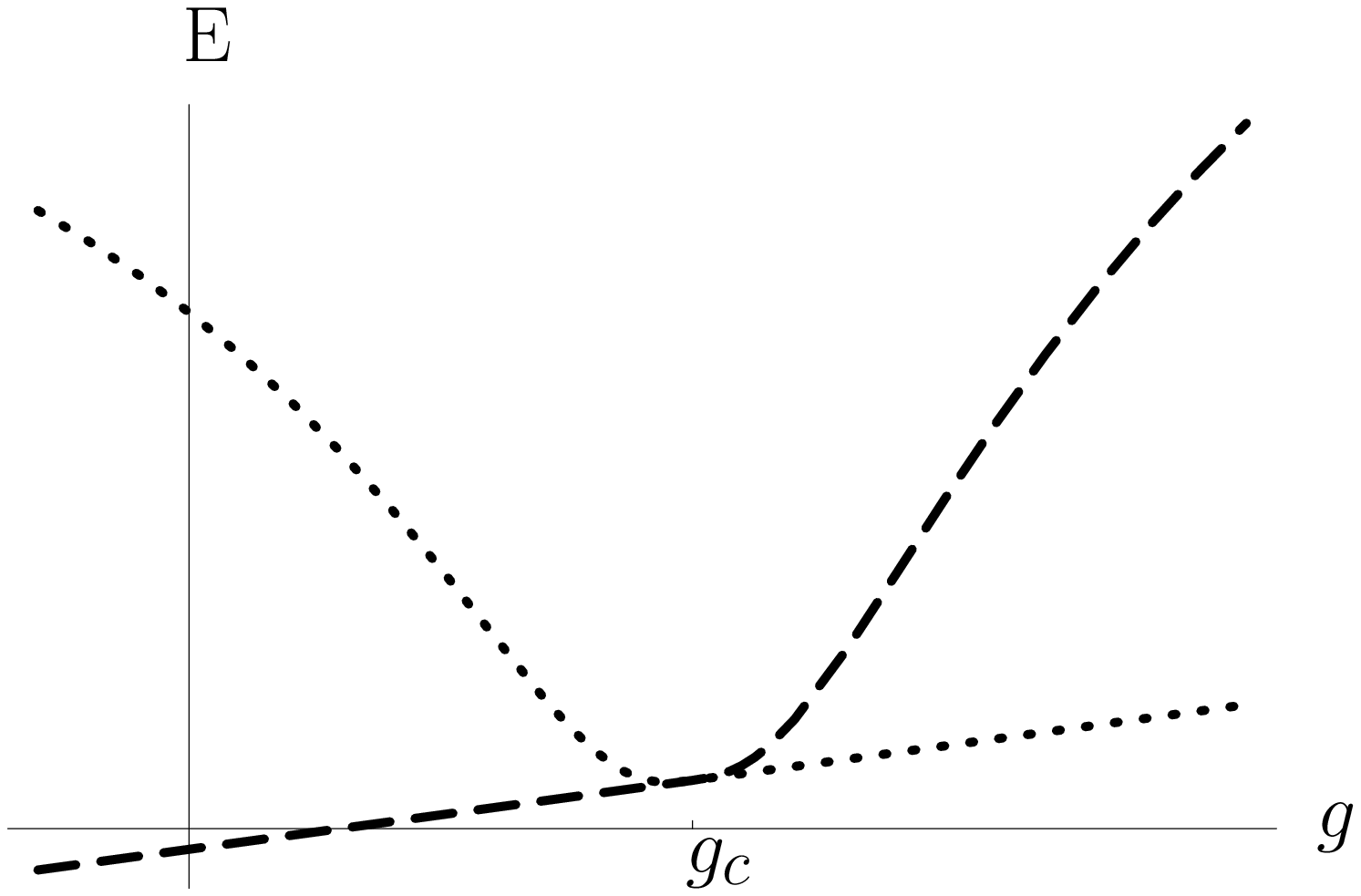,angle=0,width=0.7\linewidth}
\end{center}
\caption {Energies for the ground and first excited states as a
function of $g$. In contrast to other QPTs the ground state energy
of a MPS Hamiltonian remains (by construction) analytic at the QPT
point $g=g_c$. Nevertheless the spectral gap vanishes and the
correlation length diverges. Moreover, the non-analytic change of
the ground state at $g_c$ is reflected by an observable
non-analyticity of certain local expectation values.}
\label{figure}
\end{figure}

\emph{The Hamiltonians } --- Following the works on the AKLT model
and finitely correlated states one can always construct a local
Hamiltonian such that a given MPS is its GS: since the reduced
state density operator $\rho_k$ corresponding to $k$ sites of a
MPS has at most rank $D^2$, it has a nullspace whenever $k>\log
D^2/\log d$. Therefore, $\ket{\psi}$ is the GS of any Hamiltonian
which is a sum of (local) positive operators supported in that
nullspace. In particular, it is the GS of the Hamiltonian
\be\label{HP} H=\sum_i \tau_i(P_k)\;,\ee with $P_k$ being the
projector onto the nullspace of $\rho_k$ and $\tau_i$ its
translation to site $i$. By construction the GS energy is always
zero, i.e., it is evidently analytic in $g$.

Let us now see in which cases $H$ depends analytically on $g$ and, moreover, is such that $\ket{\psi}$ is its
unique GS for $g\neq g_c$. Consider to this end the operator
 \be
R={\cal A}^{\otimes
k}\Big(\ii_D\otimes\omega^{\otimes(k-1)}\otimes\ii_D\Big){\cal
A}^{\dag\otimes k}\;, \ee where
$\omega=\sum_{i,j=1}^D|ii\rangle\langle jj|$ and ${\cal
A}|\alpha,\beta\rangle= \sum_i [A_i]_{\alpha,\beta}|i\rangle$.
Note that, if $A_i$ depends analytically on $g$,  $R$ also depends
analytically on $g$. It is evident from the valence-bond
construction of the MPS that in general, $range(\rho_k)\subseteq
range(R)$. However, if both eigenvectors $\ket{r}$ and $\ket{l}$
have full Schmidt rank, then a straightforward calculation shows
that $range(\rho_k)= range(R)$. Hence, if this Schmidt rank
condition is satisfied on both sides of $g_c$ (which is
generically the case), then $H(g)$ indeed depends smoothly on $g$.

The uniqueness of the GS of Hamiltonians of the form (\ref{HP}) was discussed in \cite{FNW92,Nac96}, where it was
proven that it is unique if the largest eigenvalue of $E_\ii$ is non-degenerate (i.e., $g\neq g_c$),
$rank(\rho_k)=D^2$ and \be\label{EQunique} range(\rho_k)\otimes\mathbb{C}^d\cap\mathbb{C}^d\otimes
range(\rho_k)=range(\rho_{k+1})\;.\ee It is also shown there that the latter condition is always satisfied if we
replace $k$ by $k+1$, i.e., take $\label{H} H=\sum_i \tau_i(P_{k+1})\;$.

The analyticity of $H$ together with the uniqueness of its GS for $g\neq g_c$ immediately imply that a
non-analyticity in the expectation values can only be caused by a vanishing energy gap at $g_c$.

Note that a degeneracy in the GS is equivalent to a degeneracy in  $E_\ii$. This means that there is no
spontaneous symmetry breaking in one of the phases, unless $|\nu_1|=|\nu_2|$ for an entire interval (e.g., for
$g\geq g_c$). However, for a degenerate $E_\ii$ arbitrary broken discrete symmetries are
possible \cite{Nac96}.\vspace{3pt}

\emph{Examples} ---  After having discussed the general properties of MPS QPTs, we will consider some examples in
more detail. Since the MPS have a very efficient parametrization in terms of the matrices $A_i$ we can, by
imposing constraints on these matrices, \emph{engineer} systems having desired properties (symmetries, orders,
discrete symmetry breaking, etc.).\vspace*{3pt}

\emph{1. Three-body interactions:} We start by considering the case $D=d=2$, i.e., $A_1, A_2$ being two-by-two
matrices. By the arguments above every such state has a parent Hamiltonian with local three-body interactions. In
fact, many of these Hamiltonians are similar to those appearing in (triangular) optical lattices
\cite{PP04,otherHs}. We construct an example with $\mathbb{Z}_2$ symmetry by imposing the existence of a
similarity transformation which interchanges $A_1$ and $A_2$, i.e., $X^{-1}A_1X=A_2$ and $X^{-1}A_2X=A_1$. This
is indeed the case if we choose \be \hspace*{-0.2cm} A_1=\left(
\begin{array}{cc}
  0& 0 \\
  1 & 1 \\
\end{array}
\right),\  A_2=\left(
\begin{array}{cc}
  1 & g \\
  0 & 0 \\
\end{array}
\right), \mbox{ and } X = \left(
\begin{array}{cc}
  0& g \\
  1 & 0 \\
\end{array}
\right) .\ee The corresponding transfer operator $E_{\ii}$ has largest eigenvalues $1\pm g$ leading to a
singularity at $g=g_c=0$. A straightforward calculation shows a discontinuity in the first derivative of
$\langle\sigma^\alpha_i\sigma^\alpha_{i+1}\rangle (\alpha=x,y,z)$, whereas all two-point expectation values are
continuous. The magnetization in the $x$-direction can serve as an \emph{order-parameter} since $\langle
\sigma_x\rangle =4g/(1+g)^2$ for $g>0$, whereas it vanishes for $g<0$.

At the QPT point $g_c$ the state is a GHZ state.
For $g=-1$ (disordered phase) it is equal to the cluster state \cite{cluster}, and for $g=1$ (ordered phase) all
spins point in the $x$-direction.

The parent Hamiltonian is (by construction) $\mathbb{Z}_2$ symmetric: \bea H &=& \sum_i 2(g^{2}-1)\;
\sigma^z_i\sigma^z_{i+1} -
(1+g)^2\; \sigma^x_i\label{isingterm}\\
&&+ \;(g-1)^2\; \sigma^z_i\sigma^x_{i+1}\sigma^z_{i+2}\label{clusterterm}. \eea Hence, it is a combination of an
Ising interaction with transverse magnetic field (\ref{isingterm}), and a cluster state Hamiltonian
(\ref{clusterterm}). Since a constant term was omitted, the GS energy density is $e_0=-2(1+g^2)$, and one can
readily check condition (\ref{EQunique}) implying that the GS is indeed unique for $g\neq g_c=0$.

The above Hamiltonian can be embedded into a two-parameter family
\bea H(\gamma,h) &=&-\frac12\sum_i \frac{1+\gamma}2\;
\sigma_i^x-\frac{1-\gamma}2\;
\sigma_{i-1}^z\sigma_i^x\sigma_{i+1}^z\nonumber\\
&&\quad\quad\quad +h\; \sigma_i^z\sigma_{i+1}^z \nonumber\;,\eea which can be mapped onto a system of
non-interacting Fermions by a standard Jordan-Wigner transformation. Moreover, $H(\gamma,h)$ can be mapped onto
the XY-model via a duality transformation \cite{Peschel,EK}. This in turn exhibits 2$^{\rm nd}$ order QPTs on the
lines $h=\pm 1$ and $\gamma=0$ for $h\in (-1,1)$. The path parameterized by $g$ is given by $\gamma^2+h^2=1$ (the
{\it disorder-line} in the XY-model). Hence, the MPS transition occurs at the triple point of
`\emph{conventional}' QPTs exhibiting algebraically decaying correlations and diverging entanglement entropies
\cite{Peschel}. \vspace{3pt}

\emph{2. Two-body interactions:} The previous example corresponded to the case of a {\it local} order parameter.
Let us finally discuss an example with non-local \emph{string order}. To this end, consider $D=2, d=3$, i.e.,
states of a spin one chain which are GSs of nearest-neighbor interactions. The most popular MPS in this class is
certainly the GS of the spin-1 AKLT model, which exhibits a hidden (string) order. In fact, this state can be
embedded into a one-parameter family with MPS QPT. If we choose $
\big\{A_i\big\}=\big\{-\sigma_z,\sigma^-,g\sigma^+ \big\}\;, $ then the transfer operator $E_\ii$ has eigenvalues
$-1$ and $1\pm g^2$ leading to a diverging correlation length for $g\rightarrow g_c=0$. Moreover, the first
derivative of $\langle S^z\rangle$ has a discontinuity at $g_c=0$, where $\langle S^z\rangle\rightarrow 1$, i.e.,
the state becomes ferromagnetic. For $g=2$ we get the AKLT state and for $g=-2$ a state, which is equivalent to
the latter up to a local unitary transformation. For $g\rightarrow\pm\infty$ the GS becomes the N\'eel GHZ state
$(|\uparrow\downarrow\uparrow\ldots\rangle +|\downarrow\uparrow\downarrow\ldots\rangle)/\sqrt{2}$. As already
shown in \cite{FNW92} (with a different parametrization and for $g<0$) the corresponding Hamiltonian is
rotationally symmetric in the XY-plane, gapped with a non-degenerate GS (unless $g=g_c$) and has the form \bea H
&=& \sum_i (2+g^2)
\vec{S}_i\vec{S}_{i+1} + 2 \big(\vec{S}_i\vec{S}_{i+1}\big)^2 \\
&& \quad +
2(4-g^2)(S_i^z)^2+(g+2)^2(S_i^z S_{i+1}^z)^2\nonumber\\
&&\quad\nonumber +g(g+2) \big\{S_i^zS_{i+1}^z, \vec{S}_i\vec{S}_{i+1}\big\}_+\;. \eea Note that since
$E_\ii(g)=E_\ii(-g)$ the GSs corresponding to $\pm g$ merely differ by local unitaries.

\emph{Conclusion} --- Matrix product states provide the perfect
playground for investigating novel types of quantum phase
transitions that do not fit in the traditional  framework. We
provided an example of such a QPT at the triple point of
convential phase transitions, and it is easy to construct  many
more such transitions: given two predetermined MPS and associated
Hamiltonians for which they are the unique ground states, we can
always construct a one-parameter family of MPS interpolating
between the two. If on the chosen path the transfer operator
exhibits a level crossing in the largest eigenvalue, then the
system undergoes a QPT.

\emph{Acknowledgements:} The authors are grateful to the Benasque Center for Science, where parts of this work
were developed, and thank D. Perez-Garcia for interesting discussions. We acknowledge support by the DFG (SFB
631).

\vspace{-4pt}

\begin{thebibliography}{99}
\bibitem{Sac99} S. Sachdev, {\it Quantum Phase Transitions}
(Cambridge Univ. Press, Cambridge, 1999).
\bibitem{SRIS01} Q. Si, S. Rabello, K.
Ingersent, J.L. Smith, Nature {\bf 413}, 804 (2001); T. Senthil, A. Vishwanath, L. Balents, S. Sachdev, M.P.A.
Fisher, Science {\bf 303}, 1490 (2004); T. Senthil, cond-mat/0411275 (2004); T. Senthil, L. Balents, S. Sachdev,
A. Vishwanath, M.P.A. Fisher, Phys. Rev. B {\bf 70}, 144407 (2004).
\bibitem{AKLT88} I. Affleck, T. Kennedy, E.H. Lieb, H. Tasaki,
Commun. Math. Phys. {\bf 115}, 477 (1988); I. Affleck, E.H. Lieb, T. Kennedy, H. Tasaki, Phys. Rev. Lett. {\bf
59}, 799 (1987).
\bibitem{FNW92} M. Fannes, B. Nachtergaele, R.F. Werner, Commun.
Math. Phys. {\bf 144}, 443 (1992).
\bibitem{MPS} A. Kl\"umper, A. Schadschneider, J. Zittartz,
J. Phys. A {\bf 24}, L955 (1991); Z. Phys. B {\bf 87}, 281 (1992).
\bibitem{VPC04} F. Verstraete, D. Porras, J.I. Cirac,
Phys. Rev. Lett. {\bf 93}, 227205 (2004).
\bibitem{VGC04} F. Verstraete, J.J. Garcia-Ripoll, J.I. Cirac
Phys. Rev. Lett. {\bf 93}, 207204 (2004).
\bibitem{ZV04} M. Zwolak, G. Vidal,
Phys. Rev. Lett. {\bf 93}, 207205 (2004); G. Vidal, Phys. Rev. Lett. {\bf 93}, 040502 (2004).
\bibitem{VC05} F. Verstraete, J.I. Cirac, cond-mat/0505140
(2005); T. J. Osborne, quant-ph/0508031; M. B. Hastings,
cond-mat/0508554.
\bibitem{VCLRW05} F. Verstraete, J.I. Cirac, J.I. Latorre, E. Rico, M.M.
Wolf, Phys. Rev. Lett. {\bf 94}, 140601 (2005).
\bibitem{SSVCW05} C. Sch\"on, E. Solano, F. Verstraete, J.I. Cirac, M.M.
Wolf, Phys. Rev. Lett. {\bf 95}, 110503 (2005)
\bibitem{MPS2} A. Kl\"umper, A. Schadschneider, J. Zittartz,
Europhys. Lett. {\bf 24}, 293 (1993).
\bibitem{Latorre} G. Vidal, J.I. Latorre, E. Rico, A. Kitaev,
Phys. Rev. Lett. {\bf 90}, 227902 (2003); G. Refael, J.E. Moore Phys. Rev. Lett. {\bf 93}, 260602 (2004); J.P.
Keating, F. Mezzadri, Phys. Rev. Lett. {\bf 94}, 050501 (2005); V. Korepin, Phys. Rev. Lett. {\bf 92}, 096402
(2004).
\bibitem{area}M.B. Plenio, J. Eisert, J. Dreissig, M.
Cramer, Phys. Rev. Lett. {\bf 94}, 060503 (2005); M.M. Wolf, quant-ph/0503219 (2005).
\bibitem{Nac96} B. Nachtergaele, Comm. Math. Phys. {\bf 175}, 565
(1996).
\bibitem{elsewhere} F. Verstraete, M.M. Wolf, D. Perez-Garcia, J. I. Cirac,
t.b.p.
\bibitem{CC} P. Calabrese, J. Cardy, J. Stat. Mech. P06002
 (2004).
\bibitem{PEPS} F. Verstraete, J.I. Cirac, cond-mat/0407066 (2004).
\bibitem{Kac} M. Kac,
in {\it Fundamental Problems in Statistical Mechanics II}, ed. E.
G. D. Cohen (North-Holland, Amsterdam, 1968); J.A. Cuesta and A.
Sanchez, J. Stat. Phys. {\bf 115}, 869 (2004).
\bibitem{2D} H. Niggemann, A. Kl\"umper, J. Zittartz, Z. Phys. B
{\bf 104}, 103 (1997); H. Niggemann, A. Kl\"umper, J. Zittartz,
Eur. Phys. J. B {\bf 13}, 15 (2000).
\bibitem{PP04} J.K. Pachos, M.B. Plenio, Phys. Rev. Lett. {\bf 93}, 056402
(2004); J.K. Pachos, E. Rico, Phys. Rev. A {\bf 70}, 053620 (2004).
\bibitem{otherHs} Many other three-body examples can be found. For instance
$A_1=\sigma^x$, $A_2=\sqrt{g}(\ii-\sigma^z)/2$ or $A_1=\sigma^+$, $A_2=\sigma^-+\sqrt{g}(\ii+\sigma^z)/2$.
The second example leads, for instance, to \vspace{-5pt} \bea H&=& -\sum_i
g\Big(\sigma_i^x+\sigma_i^x\sigma_{i+1}^z+\sigma_i^z\sigma_{i+1}^x+\sigma_i^z\sigma_{i+1}^x\sigma_{i+2}^z\Big)\nonumber\\
&&\qquad + (1+2g^2)\sigma_i^z - 2\sigma_i^z\sigma_{i+1}^z-\sigma_i^z\sigma_{i+1}^z\sigma_{i+2}^z\;.\nonumber\eea
\bibitem{cluster} R. Raussendorf, H.J. Briegel,
Phys. Rev. Lett. {\bf 86}, 5188 (2001); Phys. Rev. Lett. {\bf 86}, 910 (2001).
\bibitem{Peschel} I. Peschel, J. Stat. Mech. P12005 (2004).
\bibitem{EK} The duality transform is $\sigma_j^x\mapsto\sigma_{j-1}^x\sigma_j^x$, $\sigma_j^z\sigma_{j+1}^z\mapsto\sigma_j^z$.\vspace*{-10pt}

\end{thebibliography}
\end{document}